 \documentclass[aps,prl,twocolumn,epsfig,floats]{revtex4}
\usepackage{graphicx,epsf,natbib,amsmath,latexsym,amssymb}
\def\be{\begin{equation}}
\def\ee{\end{equation}}
\def\bea{\begin{eqnarray}}
\def\eea{\end{eqnarray}}

\usepackage{epsfig}

\begin{document}
\author{Huazhou Wei, Sung-Po Chao, Vivek Aji}
\affiliation{Department of Physics 
and Astronomy, University of California, Riverside, CA 92521}
\title{Effect of inplane electric field on magnetotransport in helical metal}
\date{\today}
\begin{abstract}
The existence of helical surface states in a bulk insulator, with anomalous magneto-electric properties, is a remarkable new development in solid state physics. However clear signatures in transport measurements are lacking. In this paper we report on novel phenomena which provide a route to detecting massless Dirac fermions. The linear dispersion of the the fermions leads to a form of Lorentz invariance, with the Fermi velocity playing the role of the velocity of light ($c$). In a crossed electromagnetic field, the single particle states form Landau levels, whose wave-functions and energies depend on the applied \emph{in-plane electric} field but not their degeneracies. We predict the existence of oscillations of conductivity as a function of both the chemical potentials and applied electric field. This is an unique signature of the linear dispersing states due to an \emph{even-odd} effect where the conductivities of the odd Landau levels are larger than the neighboring even ones. The electric field dependence also leads to testable signatures in steady state thermopower of the non-trivial states.  
\end{abstract}
\pacs{72.15.Gd,73.43.Qt,73.50.Jt}
\maketitle

Helical Dirac fermions, massless relativistic charged particles with spin locked to their linear momentum, exist on the surface of three dimensional topological insulators (TI)\cite{Kane,Roy} and experimentally observed\cite{Konig} by angle resolved photoemission spectroscopy. Novel phenomena such as the existence of Majorana fermions in the presence of superconductor-ferromagnet interface on the surface\cite{Fu}, and a realization of a magnetic switch by tuning the conductivity with a proximate ferromagnetic film\cite{Mondal}, have been conjectured to be supported in these systems. The strong coupling of spin and orbital degrees of freedom is the source of these proposals. Magneto-transport measurements have provided some support for the existence of two dimensional conduction channels originating from the surface states\cite{Ong, Analytis}. Unfortunately, the transport measurements of TI have been masked by the conductivity of the bulk, leading to substantial difficulties in resolving the properties of the surface. 

 In this paper we report on the response of Dirac fermions under crossed electric and magnetic fields. Unlike massive fermions, the conductivity, in a quantizing magnetic field, has an oscillatory dependence on the applied electric field, wherein the even Landau levels have lower conductivity than their neighboring odd ones. This behavior is a consequence of the electric field dependence of the energies and wave-function of the Landau levels that arise due to the Lorentz invariance, allowing one to transform away the electric field in favor of an effective magnetic field. The same property allows for novel features in the steady state thermopower unique to the linear dispersion.  

 For linear spectrum tuning the in-plane electric field leads to a collapse of the Landau levels\cite{Lukose}, (un)squeezing of the oscillator states and unusual dielectric breakdown. These phenomena are robust as long as the magnetic unit cell is much larger than the crystallographic unit cell but smaller than the system size. Here we focus on the behavior of the conductivity and thermopower of the 2D Dirac fermions, with spin orbit interactions, as function of in-plane electric field. For weak fields we use the standard linear response theory\cite{sp}, while steady state properties are computed using the density matrix approach\cite{Adams}. Our chief conclusion is the prediction of magneto-oscillations as a function of applied \emph{electric} field at fixed chemical potential while no oscillations is expected at fixed particle density. In the clean limit we find a correction to the universal values of thermopower, for chemical potential at the Landau level, whose field dependence provides a signature of the surface state. For weak spin orbit scattering, our results are also applicable to graphene. 

For an electric field applied along the surface of 3D TI and magnetic field perpendicular to the surface, the Hamiltonian is 
\begin{equation*}
H_{0} = \int d^{2}\vec{r}  \psi^{\dagger}\left(\vec{r}\right)\left[v_{F}\vec{\sigma}\cdot\vec{\Pi}-\mu I -g\mu_{B}\vec{\sigma}\cdot\vec{B}-e\mathcal{E}x\right]\psi\left(\vec{r}\right)
\end{equation*}
where $\psi\left(\vec{r}\right)$ is the two component electron wave function, $v_{F}$ is the Fermi velocity, $\mu$ is the chemical potential, $\vec{\sigma} = \left\{\sigma^{x},\sigma^{y},\sigma^{z}\right\}$ are the Pauli matrices representing spin, $I$ is a $2\times 2$ unit matrix, $\vec{\Pi} = -\imath\hbar\vec{\nabla}-e\vec{A}$ is the canonical momentum, $g$ is the gyromagnetic ratio, $\mu_{B}$ is the Bohr magneton, and $\mathcal{E}$ is the magnitude of the electric field applied along the $x$ axis. The magnetic field is pointing along z-direction, perpendicular to the surface, and we choose the gauge field $\vec{A} = \left(0, Bx, 0\right)$. The Lorentz invariance of the Dirac equation implies that we can boost along y direction to a frame where the electric field does not appear in the Hamiltonian. The transformations is 
$x' = x, y'= \gamma\left(y+\beta v_{F}t\right), t' = \gamma(t+{\beta y\over{v_{F}}})$ in spatial coordinate,
$\mathcal{E}' = \gamma(\mathcal{E} - \beta v_F B), B' = \gamma(B-\beta\mathcal{E}/v_F)$ in fields, and
$\psi^{'}( \vec{r}^{'}) =  \exp[\sigma^{y}\tanh^{-1}(\beta)/2]\psi (\vec{r})$ in wavefunctions where $\gamma =1/ \sqrt{1-\beta^{2}}$ and $\beta = \mathcal{E}/v_{F}B$. In the boosted frame we solve for the eigenstates and transform back to get the exact eigenstates in finite electric field in the lab frame. These transformations are valid as long as $\beta \leqslant 1$. For $\beta > 1$ the transformations lead to scattering states. The transport properties for this case are beyond the scope of this investigation.
In the lab frame the energy eigenvalues are
\begin{eqnarray}\nonumber
E_{n} &=& sgn(n){\hbar v_{F}\over{\gamma^{3/2} l_B}}\sqrt{2\left|n\right| + \kappa_{0}\gamma B}-\beta\hbar v_{F}k_{y}, n \neq 0 \\ 
&=& -\gamma^{-1}g\mu_{B}B -\beta\hbar v_{F}k_{y} \hspace{0.5cm}, n = 0 
\end{eqnarray}
with $l_B=\sqrt{\hbar/eB}$ as the magnetic length, $\kappa_{0} = g^{2}\mu_{B}^{2}/\hbar v_{F}^{2}e$, and $n$ as Landau level index for Dirac fermions. The corresponding spatial wave functions for $n\neq 0$ in the lab frame are
\begin{eqnarray}\label{wf1}
 \psi_{n}\left(\vec{r}\right) =\frac{N_{n}e^{\imath k_{y}y}}{\sqrt{N_y}}\left(\begin{array}{c}\cosh(\frac{\theta}{2})\phi_{\left|n\right|}+\imath \alpha_{n}\sinh(\frac{\theta}{2})\phi_{\left|n\right|-1} \\-\imath\sinh(\frac{\theta}{2})\phi_{\left|n\right|}+\alpha_{n}\cosh(\frac{\theta}{2})\phi_{\left|n\right|-1}\end{array}\right)
 \end{eqnarray}
where we have used $k_y y-E t=k_y'y'-E't'$ and take the spatial part of the wavefunction. $N_y$ is the normalization factor coming from periodic boundary condition in the $y$ direction. We denote the parameters in the lab frame by $N_{n}= {1/{\sqrt{1+\left|\alpha_{n}\right|^{2}}}}$, $\tanh(\theta)=\beta$, $\alpha_{n}=-\imath\left(sgn(n)\sqrt{1+{\gamma c^{2}B^{2}/{\left|n\right|}}}+\sqrt{\gamma}cB/\sqrt{\left| n\right|}\right)$, 
$c= l_{B} g\mu_{B}/\sqrt{2}\hbar v_{F}$, and $\phi_{|n|}$ is given by
$
\phi_{|n|} = {1\over \gamma^{3/4}}\sqrt{1\over{\sqrt{\pi}2^{|n|}|n|!l_{B}}}H_{|n|}\left(\xi_{n}\right)e^{-{\xi_{n}^{2}\over{2}}}
$. Here $H_{|n|}$ is the $|n|$th order Hermite polynomial, $\xi_{n} = \left(x - l_{B}^{2}k_{y}-l_{n}\right)/\sqrt{\gamma}l_{B}$ with $l_{n} = sgn(n)\beta l_{B}\sqrt{\gamma}\sqrt{2\left|n\right|+\kappa_{0}\gamma B}$, and $l_{0} = -\beta l_{B}\sqrt{\kappa_{0}\gamma^2 B}$. For $n=0$
\begin{eqnarray}
 \psi_{0}\left(\vec{r}\right) = \left(\begin{array}{c}\cosh(\theta/2) \\-\imath\sinh(\theta/2)\end{array}\right)\phi_{0}\frac{e^{\imath k_{y}y}}{\sqrt{N_y}}
\end{eqnarray}
In the lab frame each component is a linear combination of Hermite polynomials of different index $n$. Moreover $\beta$, being a function of the electric field, varies from zero to $1$. Since $\gamma$ diverges as $\beta$ approaches $1$, the Landau levels will collapse\cite{Lukose}. The fact that the relative strength of the orbital and Zeeman terms can be manipulated by the electric field or the magnetic field allows richer dependence of transport coefficients in the surface state of topological insulators. To compute conductivity we use the density matrix approach\cite{Adams} to obtain the charge current.
The electrical conductivity tensor $\sigma_{\alpha\beta}$ is given by $J_{\alpha} = \sigma_{\alpha\beta}E_{\beta}$. For an ensemble described by the density matrix $\rho$, we get
\begin{equation}\label{cur}
J(\vec{r})^{i} =-e v_{F}\sum_{mn}\psi_{n}^{\dag}\sigma^{i}\psi_{m}\rho_{mn}
\end{equation}
To obtain transport coefficient we introduce a scattering potential $V_s(\vec{r})$. The density matrix satisfies
\begin{equation}
-i {\partial\rho\over{\partial t}} = [\rho, H],\hspace{0.5cm} H = H_{0}+V_s
\end{equation}
A Laplace transform yields
$-i s P(s) = [P(s), H]-i  \rho(0)$,
where $P(s)=\int_{0}^{\infty}e^{-st}\rho(t)dt$ is the Laplace transform of $\rho$. We decompose the $P(s)$ into three parts
$
P_{mn}(s) = \frac{1}{s}(f_{n}\delta_{mn} + D_{mn}(s) + G_{mn}(s))
$
where $f_n/s$ is the Laplace transform of the initial density matrix which is assumed to be the Fermi Dirac distribution $f_n=f(\omega_n)=1/(\exp[(\omega_n-\mu)/k_BT]+1)$ with $\omega_{n} = sgn(n) {\hbar v_{F}\over{\gamma^{3/2}l_{B}}}\sqrt{2\left|n\right| + \kappa_{0}B\gamma}$. $D_{mn}$ is the part of the density matrix which is nonzero only if $k_{y}$ of the states $m$ and $n$ are identical and $G_{mn}$ is nonzero only if $k_{y}$ of the states $m$ and $n$ are different. Expanding to linear order in the scattering matrix the density matrix is\cite{Adams}
$G_{mn} = {f_{mn}V_{mn}\over {\omega_{mn}-\beta\hbar v_{F}k_{ymn}-i s}}$ and 
$D_{mn} =\sum_{\nu}{{V_{m\nu}V_{\nu n}}\over{\omega_{mn}-i s}}\Big[{f_{m\nu}\over {\omega_{m\nu}-\beta\hbar v_{F}k_{ym\nu}-i s}}-{f_{\nu n}\over {\omega_{\nu n}-\beta\hbar v_{F}k_{y\nu n}-i s}}\Big]$.
Here $V_{nm}$ is the matrix element of the scattering potential, $\omega_{mn} = \omega_{m}-\omega_{n}$, $k_{yn\nu} = k_{yn}-k_{y\nu}$ and $f_{\mu\nu} = f_{\mu}-f_{\nu}$. For an arbitrary potential we write its Fourier component $V_s(\vec{r})=\sum_{\vec{q}}V_{\vec{q}}e^{i\vec{q}\cdot\vec{r}}$ and the matrix element $V_{nm}=\int d\vec{r}\psi^{\dagger}_n(\vec{r})V(\vec{r})\psi_m(\vec{r})=\sum_{\vec{q}}V_{\vec{q}nm}$. In this paper we study the case of random impurity potential where $V_s(\vec{r})=\sum_i V_0 \delta(\vec{r}-\vec{r}_i)$. For a random array of scatterers, the phase factors appearing in $G_{nm}$ will average to zero. We only need to consider matrix elements in the current operator where the two states have the same $k_{y}$.

 We compute the steady state current by taking the long time limit or steady state density matrix $\rho(t\rightarrow\infty)$. This is obtained by taking Laplace transform parameter $s$ through positive zero as $\rho_{mn}(t\rightarrow\infty)=\lim_{s\rightarrow 0^+}sP_{mn}(s)$. For off diagonal elements of the density matrix only $D_{n m}$ is relevant. This dissipation term is controlled by the imaginary part and the expression is
\begin{eqnarray}\nonumber
&&D_{n m} = {\pi \imath\over {\omega_{n m}}}\sum_{p}V_{np}V_{pm}\Big[f_{np}\delta\left(\omega_{np}-\beta\hbar v_{F}k_{npy}\right)\\\label{dmn}&&+f_{mp}\delta\left(\omega_{mp}-\beta\hbar v_{F}k_{mpy}\right)\Big]
\end{eqnarray}

\noindent where $k_{nmy}$ is the difference in value of $k_{y}$ of the states $n$ and $m$. Take $\beta\ll 1$ in Eq.(\ref{cur}) and Eq.(\ref{dmn}) we obtain the diagonal conductivity $\sigma_{xx}=J_x/E$ in the linear response regime as
\begin{eqnarray}\label{lin}
\sigma_{xx}=\frac{e^2}{\hbar}n_i\sum_{n,\nu}2\pi|V_{n\nu}|^2(k_{n\nu y}l_B)^2\frac{\partial f}{\partial \omega}\Big|_{\omega=\omega_n}\delta(\omega_{n\nu})
\end{eqnarray}
which is the same as the conductivity derived from lowest order Kubo formula with $n_i$ denoting impurity density. In Eq.(\ref{lin}) we have used lowering operators on $V_{n+1,\nu}$ to connect it with $V_{n\nu}$ as done in Ref.\cite{Adams} for 2DEGs case. For large in-plane field we compute numerically the diagonal conductivity as a function of chemical potential in different in-plane electric field for both conventional 2DEGs and Dirac fermions. 

\begin{figure}
\centering
\begin{tabular}{cc}
\epsfig{file=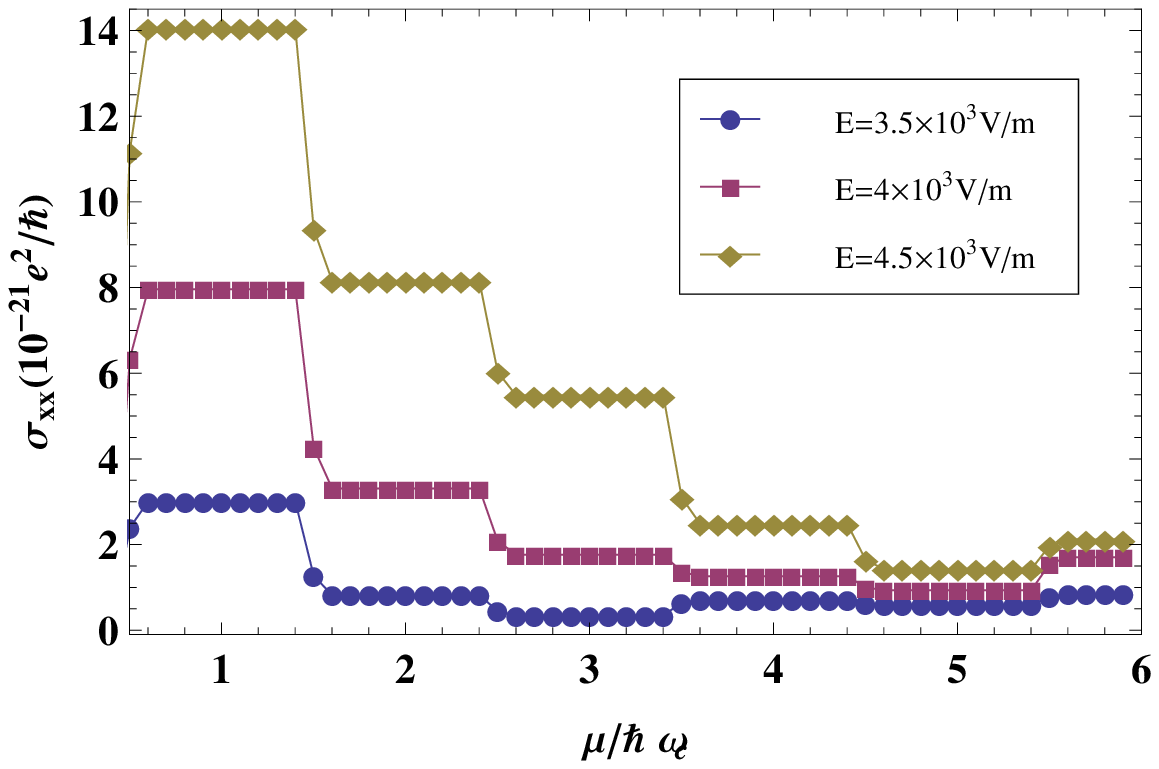,width=0.5\linewidth,clip=} & 
\epsfig{file=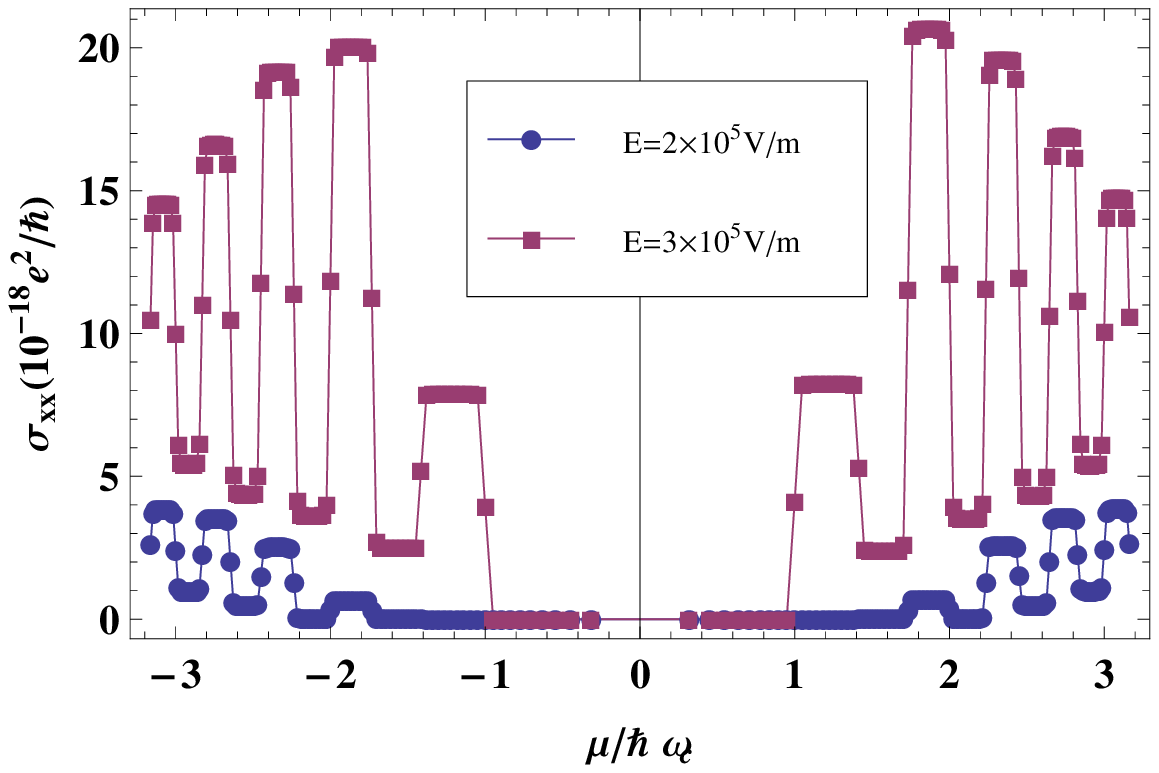,width=0.5\linewidth,clip=} \\
\epsfig{file=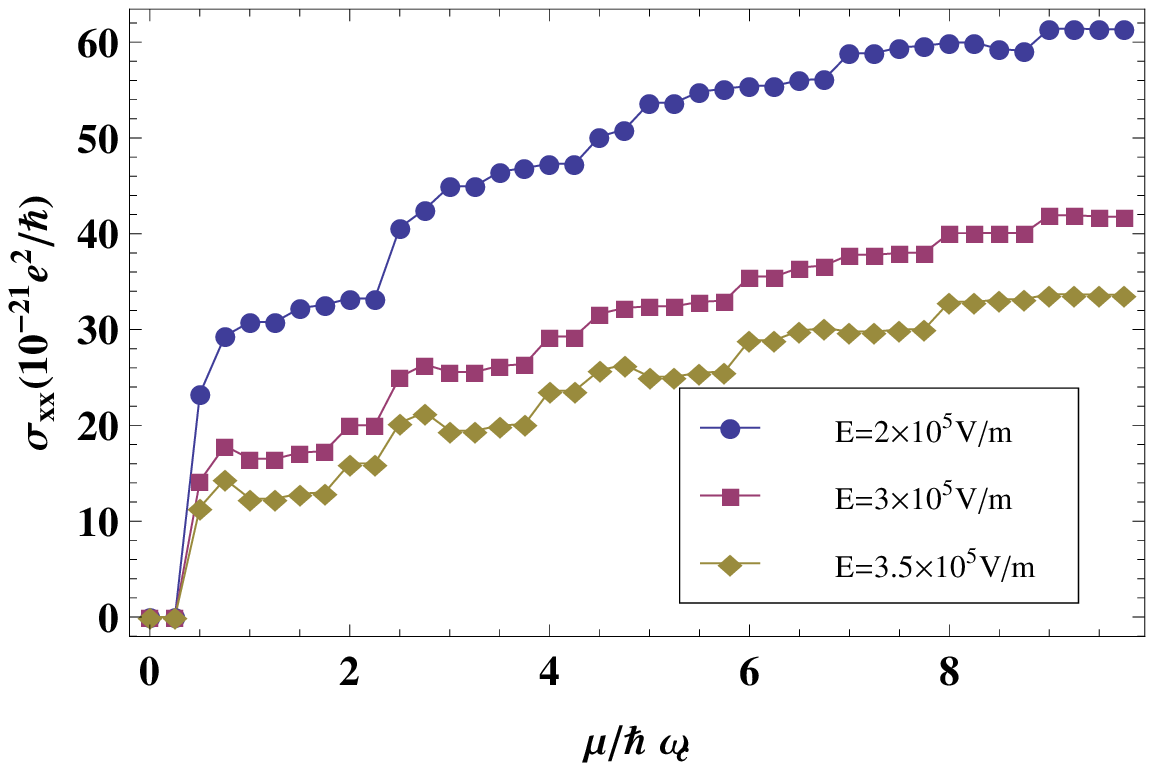,width=0.5\linewidth,clip=} &
\epsfig{file=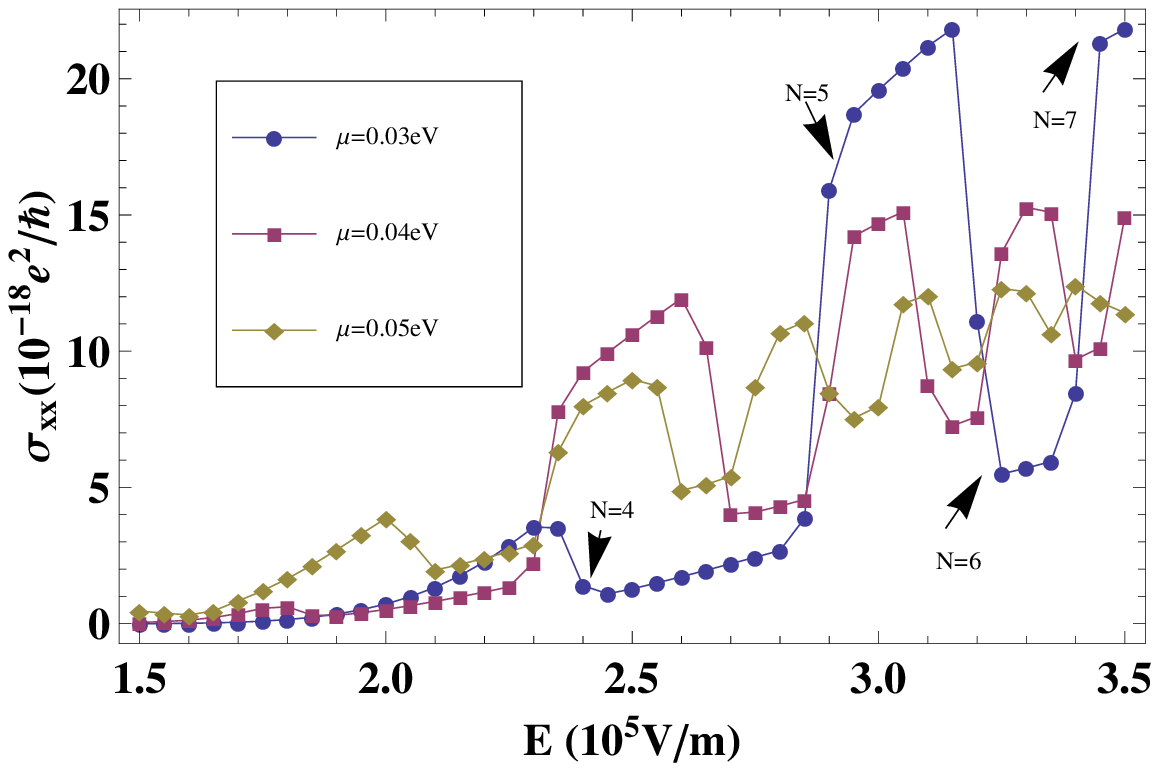,width=0.5\linewidth,clip=} \\
\epsfig{file=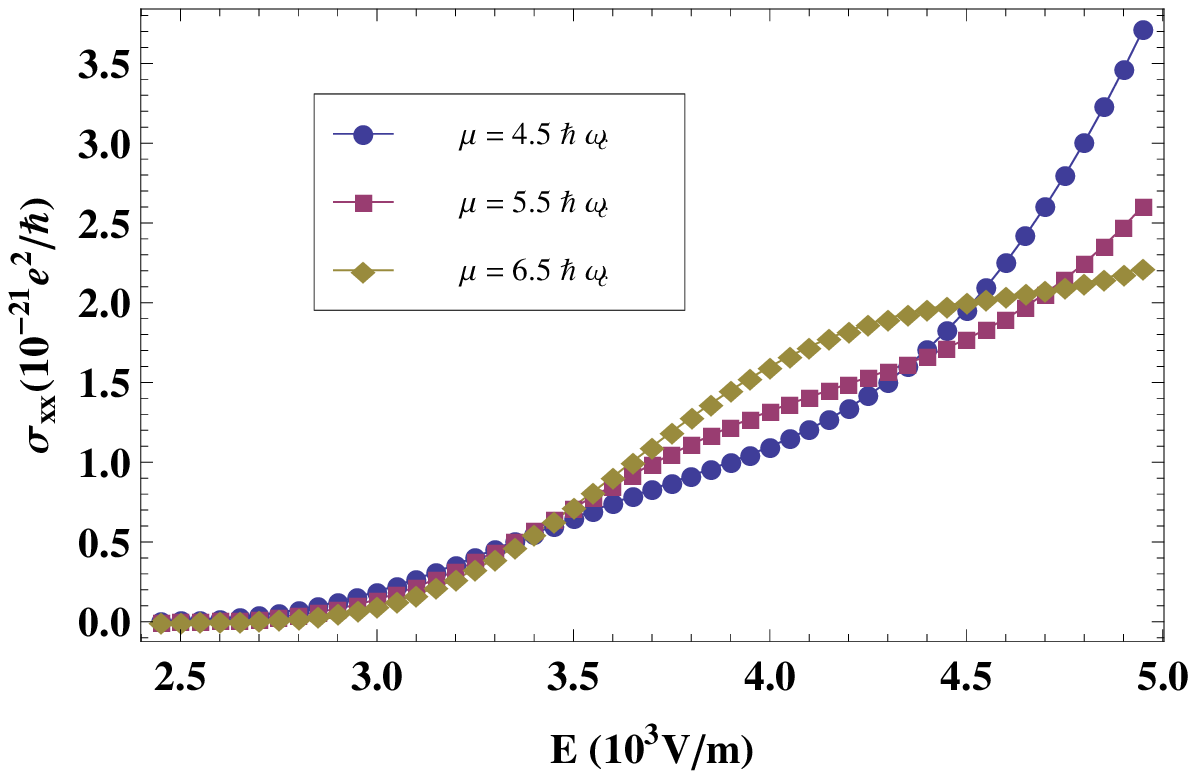,width=0.5\linewidth,clip=} &
\epsfig{file=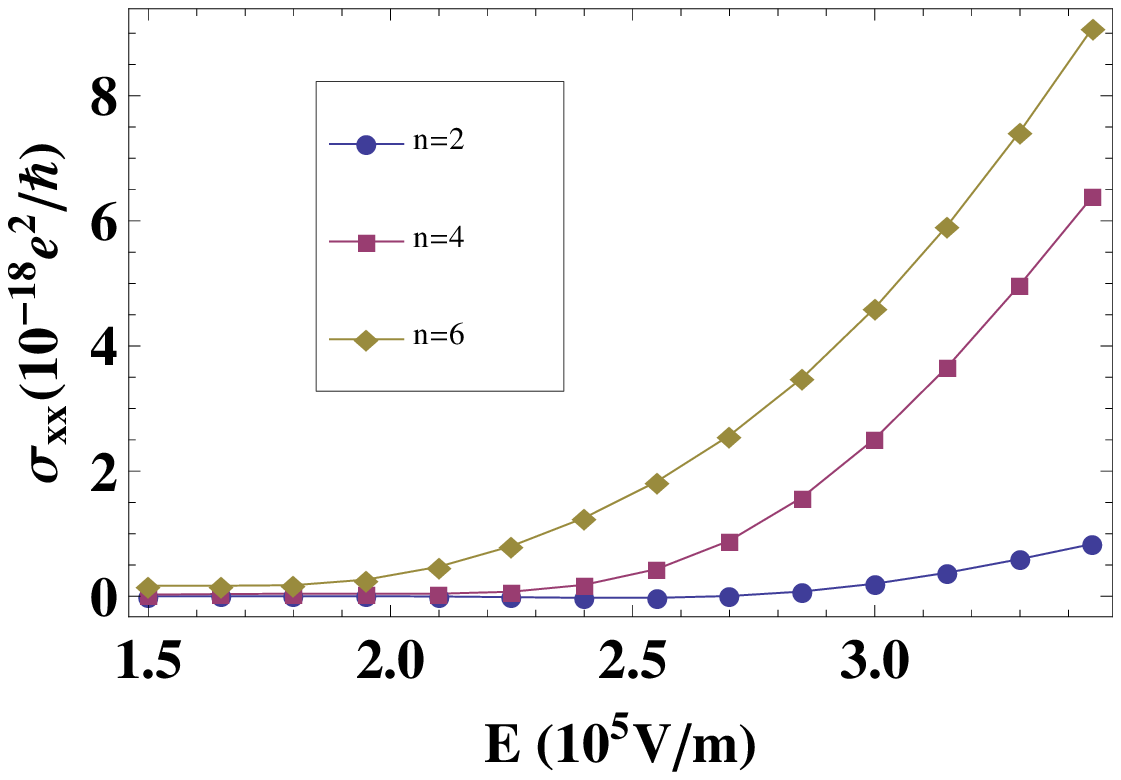,width=0.5\linewidth,clip=} 
\end{tabular}

\caption{(Color online) Top: $\sigma_{xx}$ v.s. $\mu/\hbar\omega_c$. Top right: the  Dirac fermions with $E/v_F B=0.4$ (blue circle) and $0.6$ (purple square). Top left: 2DEGs with $eEl_B/\hbar\omega_c\simeq 0.77$ (blue circle), $0.88$ (purple square), and $ 0.99$ (brown diamond).  Middle right: $\sigma_{xx}$ v.s. $E$ for chemical potential $\mu\simeq1.7/\hbar\omega_c$ (blue circle), $2.2$ (purple square), and $2.8$ (brown diamond). The Landau level positions are shown explicitly for $\mu\simeq1.7/\hbar\omega_c$. Middle left: $\sigma_{xx}$ v.s. $\mu/\hbar\omega_c$ for 2DEGs with $eEl_B/\hbar\omega_c\simeq 44$ (blue circle), $66$ (purple square), and $77$ (brown diamond). Bottom right: $\sigma_{xx}$ v.s. $E(10^5 V/m)$ for $\mu=\omega_2$ (blue circle), $\mu=\omega_4$ (purple square), and $\mu=\omega_6$ (brown diamond). Bottom left: $\sigma_{xx}$ v.s. $E(10^5 V/m)$ for 2DEGs with chemical potential $\mu=4.5 \hbar\omega_c$ (blue circle), $\mu=5.5 \hbar\omega_c$ (purple square), and $\mu=6.5 \hbar\omega_c$ (brown diamond). The order of magnitude ( $10^{-18}$ and $10^{-21}$ for Dirac and 2DEGs respectively) is set by the scattering matrix element given by $\sum_{n\nu}|V_{n\nu}|^2\delta(E_n-\mu)/\hbar\omega_c$ with $V_0\simeq0.18\hbar\omega_c$ at $T=1K$ for the Dirac spectrum and $V_0\simeq0.11\hbar\omega_c$ at $T=10^{-2}K$ for the 2DEGs case.
\label{fig1}}
\end{figure}

The results for longitudinal conductivity for 2DEGs and Dirac fermions are shown in the left and right panel of fig.\ref{fig1} respectively. Within our approximation the Landau levels have zero width and the conductivities show steps as a function of chemical potential (see the top left and right, and the middle left panels of fig.\ref{fig1}). Note that all features that we discuss have sharp transitions across Landau levels. This is related to the lack of impurity and thermal broadening. The results are valid as long as the broadening is less than the Landau level spacing. For 2DEGs the conductivity either increases or decreases as a function of carrier density depending on whether $e E l_B\over \hbar\omega_c$ is $\gg 1$ or $\lesssim 1$. In the former case the increase of available phase space wins out against the decrease in the matrix elements $V_{np}$ as a function of chemical potential, while for the latter the opposite is true. The non-monotonic behavior is also expected as a function of electric field strength as seen in the bottom left panel. 

The behavior of the Dirac spectrum is far richer as there are a number of competing effects. The Fermi function restricts contribution to the density matrices from levels near the chemical potential. Since the levels get closer for large $n$, the energy denominator in Eq.(\ref{dmn}) decreases, leading to an increase in the density matrix.  The phase space satisfying the delta functions also increases due to the decrease in the level spacing at large $n$ further enhancing $D_{nm}$. On the other hand the matrix elements, $V_{np}$, decrease for large chemical potentials. To understand why, note that the two states $n$ and $p$ differ in their $k_{y}$ values. Consequently their centers are separated in real space. As the chemical potential increases, the difference in $k_{y}$ values and hence the distance in real space, goes down. The two states are orthogonal, if they had the same center and thus the decrease at large chemical potentials. There are two further competing effects in the presence of an electric field. An increase in the distance between the centers due to the change in $l_{n}$ competes with the decrease in the difference in $k_{y}$ due to the Landau level collapse. The net result of all these variations is an $even$-$odd$ effect, where the conductivity of the odd Landau levels are larger than their neighboring even levels as seen in the top right panel of fig.\ref{fig1}\cite{note}. At large chemical potentials, the effect is washed out as the scattering process connects multiple levels with different symmetries.

As evident in fig.\ref{fig1} and Eq.(\ref{dmn}), the order of magnitude of the conductivity is set by the impurity potential and impurity density within the perturbative approach employed here. To get to conductivities of $e^{2}/h$, a larger value of the impurity potential is required, beyond the scope of this calculation. Nevertheless, the Landau level collapse and the even-odd effect are more general properties arising from the linear dispersion. To what extent these results will be modified within a fully self-consistent treatment is a subject of future investigation.

For Dirac fermions the conductivity also shows oscillatory behavior for fixed chemical potential as shown in middle right of fig.\ref{fig1}. This quantum oscillation has to do with the Landau level collapsing\cite{Lukose} with increasing field mainly due to the $\gamma^{-3/2}$ factor in $\omega_n$. The $even$-$odd$ effect translates to an oscillation as opposed to a series of plateaus. For fixed charge density, as implemented in the experiment by fixing the gate voltage, no oscillations are expected (bottom right panel of fig.\ref{fig1}) as the degeneracy of the Landau level is independent of the applied electric field.

\begin{figure}[t]
\includegraphics[width=0.7\columnwidth, clip]{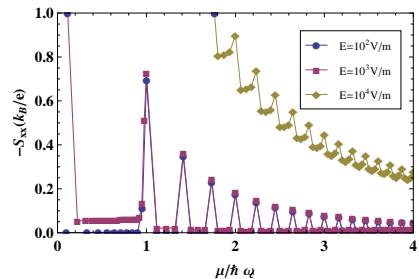}
\caption{(Color online) Negative thermopower $-S_{xx}(k_B/e)$ v.s. chemical potential $\mu/\hbar\omega_c$ for the clean 2D Dirac fermions. Temperature is fixed at $10^{-2}\hbar\omega_c/k_B$. $g=2$ and $E=10^2 V/m$ for blue circle line, $E=10^3 V/m$ for purple square line, $E=10^4 V/m$ for brown diamond line.} 
	  \label{fig2}
\end{figure}

Another interesting measurable quantity is the steady state thermopower. Consider a system with finite current due to an applied voltage $V$. We impose small source-drain temperature difference, such that the temperature gradient can be treated in perturbation, with a mean temperature $T$. The particle current density $J_i^P$ is\cite{Jonson}
$
J_i^P=L^{11}_{ij}\frac{e}{T}\partial_j V+L_{ij}^{12}\partial_j \frac{1}{T} =J_i^{P(1)}+J_i^{P(2)}
$
with $J_i^{P(1)} = -J^{i}/e$ where $J^{i}$ is the electric current, and $J_i^{P(2)}$ is the particle current coming from temperature gradient. The electric conductivity $\sigma_{ij}=-(e^2/T)L_{ij}^{11}$. Note that $L_{ij}^{11}$ is $\emph{not}$ given by the Kubo formula as $-\nabla V$ is not small enough to be in the linear response regime. We define the out of equilibrium steady state thermopower $S_{ij}$ as the response to a small temperature gradient and small perturbation $\nabla V'$ around the applied $\nabla V$. The out of equilibrium thermopower
is
\begin{eqnarray}\label{13}
 S_{ij}=\frac{-1}{eT}(L^{11})^{-1}_{ik}L_{kj}^{12}.
\end{eqnarray}
Since the temperature gradient is treated within linear response and the large uniform electric field does not break time reversal symmetry, the $L_{ij}^{12}$ is 
$L_{ij}^{12}=\bar{L}_{ij}^{12}+P_{ij}$
with $\bar{L}_{ij}^{12}=\lim_{\omega\rightarrow 0}\frac{-T}{\omega V}\int_{0}^{\infty} dt e^{i\omega t}Tr(\rho[\hat{J}_i^P(t),\hat{J}_j^Q(0)])$ representing the Kubo term and $P_{ij}=\frac{-T}{2\rm{L_i}} Tr(\rho\{\hat{r}_i,\hat{J}_j^P\})$ the correction term due to diamagnetic currents\cite{Jonson}. 
Here $\hat{J}_i^P=v_F \hat{\psi}^{\dagger}\sigma^i\hat{\psi}$ is the particle current operator, $\hat{r}_i$ is the position operator, and $\hat{J}_i^Q=(\hat{H} \hat{J}_i^P+\hat{J}_i^P\hat{H})/2-\mu \hat{J}_i^P$ is the heat current operator. $\rho$ is the steady state density matrix. The time dependence is expressed in the Heisenberg representation. The position operator is related to the particle current operator as $\hat{J}_i^P=\frac{i}{\hbar\rm{L_i}}[\hat{H},\hat{r}_i]$. In the clean limit where impurity potential energy $V\rightarrow 0$, the diagonal conductivity satisfy $\sigma_{xx}=\sigma_{yy}=0$ even in the presence of finite electric field (with $E<v_F B$). Within this clean limit the low temperature thermopower in the finite electric field 
for 2D Dirac spectrum is shown in fig.\ref{fig2}. With increasing electric field we see that the thermopower increases and begins to deviates from its universal value when $\beta\sim 10^{-3}$ ($E \sim 10^{3} V/m$), which is much smaller than the linear response regime seen in the transport 
measurement of current as shown in the bottom right of fig.\ref{fig1}.

In conventional 2DEGs the thermopower (following the derivation in \cite{Adams}) at finite electric field, low temperature and for $\mu\simeq \hbar\omega_c (N+1/2)\equiv E_N$ is $S_{xx}\simeq -\frac{k_B \ln 2}{e(N+\frac{1}{2})}-\frac{4 k_B}{e}\frac{\frac{m}{2}(\frac{E}{B})^2}{k_B T(N+\frac{1}{2})}$ . The second term leads to an increase in the large electric field. In the linear response regime the 2DEGs has universal thermopower $S_{xx}\simeq -\frac{k_B \ln 2}{e(N+\frac{1}{2})}$ at $\mu\simeq E_N$, independent of the $B$ field strength. From Eq.(\ref{13}) the thermopower in the linear response regime for clean 2D Dirac gas with negligible Zeeman term is
$
S_{xx}=\frac{-2}{eT}\frac{\sum_n\int_{\omega_n}^{\infty}d\epsilon(\epsilon-\mu)\frac{\partial f}{\partial \epsilon}}{\sum_n\tanh\left(\frac{\omega_n-\mu}{2k_B T}\right)}
$, 
which is the same as derived by Kubo formula in \cite{sp}. At $\mu\simeq \omega_n$ the thermopower $S_{xx}\simeq -\frac{k_B \ln 2}{e n}$ is universal in the linear response regime. For chemical potential in between Landau levels the lack of density of states leads to vanishing thermopower. This is related to the lack of impurity and thermal broadening of the Landau levels. For sufficient large in-plane field the universal peaks in the thermopower are overwhelmed by the entropic contribution from the field, similar to its counterpart in the 2DEGs where the $E^2$ correction gives rise to overall increase in the thermopower. For materials with large gyromagnetic ratio or large Zeeman effect the peak positions in $S_{xx}$ depends on the strength of magnetic field and the peak height is not universal\cite{sp}. As the nonequilibrium feature depends on the ratio between electric and magnetic field, larger magnetic field results show features similar to the equilibrium system.

In summary, we use Lorentz boost to find the exact eigenstates in the presence of crossed electric and magnetic field for Dirac fermions realized on the clean surface of 3D TI. The effect of weak random impurities are considered perturbatively through density matrix formulation. In the linear response regime the conductivity obtained is the same as the lowest order Kubo formula. For general in-plane field strength we carry out the numerical computation for both Dirac fermions and 2DEGs. Quantum oscillation is observed as a function of in-plane field for fixed chemical potential due to the Landau level collapse, which serves as unique signature of Dirac fermions. In addition to the charge transport, we define the steady state thermopower to extend the thermoelectric measurement to the nonequilibrium steady state. The external field, acting as the driving source for entropy production, enhances the overall thermopower. For sufficient large field the universal peaks in the thermopower are wiped out. 

Finally we discuss the observability of the effects discussed. While the even odd effect is visible for clean samples (i.e. level broadening much less than level spacing and well defined parities of the wave-functions), a full self consistent calculation is currently being pursued to determine the regime of breakdown\cite{Gusynin,vincent,Ramer} and will be published elsewhere. Such a study will also help establish the extent of the validity of Lorentz transformation for systems with dissipation. Joule heating can potentially wipe out the effect. The net heat flux produced by a 10$\mu$m $\times$ 10$\mu$m device is 0.1mW at $E\sim 10^{5}$V/m ($\sigma\sim e^{2}/h$). As long as the cooling is efficient enough to overcome this, the electric field effects are observable. Magneto-oscillations in linear response have already been measured both in topological insulators\cite{Ong} and graphene\cite{Ong2}. The latter has low enough minimum conductivity to make these measurements feasible.

We thank Chandra Varma and Douglas MacLaughlin for helpful comments and suggestions. S-PC and VA acknowledge the financial support by University of California at Riverside through the initial complement.


\begin{thebibliography}{999}
\bibitem{note} The wavefunction overlap in the density matrix goes like $|\phi_{|n|}(x)|^2(|\phi_{|n-1|}(x-b)|^2-|\phi_{|n+1|}(x+b)|^2)$ in the lowest order expansion in $\beta$ with $b/l_B\propto \gamma^{1/2}/\beta$. For $0.1\le\beta\le0.7$ the integrated value for odd $n$ is larger than neighboring even $n$.
\end{thebibliography}
\end{document}